\documentclass[aps,prd,showpacs,twocolumn,
10pt,superscriptaddress,floatfix,nofootinbib,longbibliography]{revtex4-1}   %twocolumn,
\usepackage{amssymb,amsfonts,amsmath,bm,bbm}   
\usepackage{graphicx}% Include figure files
\usepackage{dcolumn}% Align table columns on decimal point
\usepackage{bm}% bold math
\usepackage{CJK}
\usepackage{minted}  %%%%
\usepackage[%dvipdfm,
colorlinks=true,
linkcolor=blue,%black,
breaklinks=true,
urlcolor=blue,
citecolor=blue]{hyperref}
\usepackage{bbold}
\usepackage{epstopdf}

\newcommand{\HNUST}{\affiliation{
Hunan Provincial Key Laboratory of Intelligent Sensors and Advanced Sensor Materials, \\ School of Physics and Electronics, Hunan University of Science and Technology, Xiangtan 411201, China}}

\newcommand{\IHEPparti}{\affiliation{
State Key Laboratory of Particle Astrophysics, Institute of High Energy Physics, \\Chinese Academy of Sciences, Beijing 100049, China}}

\newcommand{\IMNU}{\affiliation{
College of Physics and Electronic Information, Inner Mongolia Normal University, Hohhot, 010022, China}}

\begin{document}
%\preprint{APS/123-QED}

\title{Topological observables and domain wall tension from finite temperature chiral perturbation theory
}

\author{Zhen-Yan Lu}\email{luzhenyan@hnust.edu.cn}
\HNUST

\author{Quan Tang}
\email{tangquan@mail.hnust.edu.cn}
\HNUST

\author{Shu-Peng Wang}
\email{wsp@mail.hnust.edu.cn}
\HNUST

\author{Yang Huang} 
\email{yanghuang@mail.bnu.edu.cn}
\HNUST

\author{Zhen Zhang} \email{zhangzhen@ihep.ac.cn}\IHEPparti

\author{Bonan Zhang}\email{bnzhang@imnu.edu.cn}\IMNU

\date{\today}

\begin{abstract}
Within the framework of SU(2) chiral perturbation theory, we derive the general solution of the QCD $\theta$-vacuum for an arbitrary vacuum phase, explicitly incorporating isospin-breaking effects from the light quark mass difference, and compute the temperature dependence of the topological susceptibility, higher-order cumulants, and the domain wall tension up to next-to-leading order. We find that the topological susceptibility agrees with lattice data at low temperatures but deviates at higher temperatures as expected from the breakdown of the chiral expansion; moreover, we demonstrate that the normalized fourth-order cumulant and the domain wall tension decrease monotonically with increasing temperature, while the normalized sixth-order cumulant exhibits the opposite behavior. These results extend earlier analyses by showing how isospin breaking reshapes the full hierarchy of topological charge cumulants and the dynamics of $\theta$-vacuum domain walls, thereby offering new theoretical input on the $\theta$-vacuum properties, which are relevant for axion-related effective theories in hot QCD matter.

\end{abstract}

\maketitle
%\tableofcontents

\begin{CJK*}{UTF8}{gbsn}%{gkai}

%%%%%%%%%%%%%%%%%%%%%%%%%%%%%%%%%%%%%%%%
\section{Introduction} \label{sec:INTRODUCTION}

Quantum chromodynamics (QCD) is the fundamental theory of the strong interaction~\cite{Gross:2022hyw,Achenbach:2023pba}. One of its most remarkable features is the rich vacuum structure associated with topologically non-trivial gauge configurations~\cite{Vonk:2019kwv,Vicari:2008jw}. These configurations, characterized by the integer-valued topological charge
\begin{align} \label{eq:Qcharge}
Q=\frac{g_s^2}{32 \pi^2}\epsilon^{\mu \nu \rho \sigma}\int d^4x~ G_{\mu\nu}^c(x)G_{\rho\sigma}^c(x),
\end{align}
give rise to distinct $\theta$-vacuum sectors, connected by tunneling transitions mediated by instantons~\cite{Gross:1980br,Belavin:1975fg}. In Eq.~(\ref{eq:Qcharge}), $G_{\mu\nu}^c$ denotes the gluon field strength tensor with color index $c$, and $\epsilon^{\mu \nu \rho \sigma}$ is the totally antisymmetric Levi-Civita tensor. The $\theta$-term that couples to the topological charge has profound phenomenological implications~\cite{Ubaldi:2008nf,Acharya:2015pya,Lee:2020tmi}, including the induction of a neutron electric dipole moment. The extremely tight upper bound $|\theta|\lesssim10^{-10}$ inferred from experiments~\cite{Baker:2006ts,Griffith:2009zz,Parker:2015yka,Graner:2016ses,Guo:2015tla} underlies the strong CP problem and motivates axion physics as a resolution~\cite{DiLuzio:2020wdo,Antel:2023hkf,Choi:2020rgn}. In this context, the $\theta$-dependent observables computed in this work provide first-principles input to the axion effective theory. Identifying the QCD vacuum angle with the axion field through $\theta = a/f_a$, one can directly obtain the axion potential from the $\theta$-dependent vacuum free energy as
$ V(a)=e_\text{vac}(a/f_a)$, and $m_a^2 = \chi_t/f_a^2 $~\cite{Lu:2020rhp}. 
The higher-order cumulants $b_2$, $b_4$, and beyond quantify the non-Gaussian features of $V(\theta)$, refining the understanding of axion dynamics in the early Universe and in hot QCD matter. The temperature dependence of $\chi_t$ and these cumulants obtained in this study thus provides essential quantitative input for axion phenomenology~\cite{Marsh:2015xka,Carenza:2024ehj}.

A central quantity that encodes the $\theta$-dependence of QCD is the topological susceptibility $\chi_t$, together with higher-order cumulants $b_{2},b_{4},\dots$, which determine the non-Gaussian structure of the topological charge distribution. The dilute instanton gas approximation provides a useful picture at very high temperatures~\cite{Petreczky:2016vrs,Aarts:2023vsf}, while lattice QCD offers reliable information at vanishing temperature and at asymptotically high temperature~\cite{Borsanyi:2016ksw,Kotov:2025ilm,Frison:2016vuc,Burger:2018fvb}. However, theoretical control in the phenomenologically relevant regime of low to intermediate temperatures, especially below the chiral transition, remains limited.

Chiral perturbation theory (CHPT) provides a systematic and model-independent effective field theory framework for describing the low-energy dynamics of QCD~\cite{Gasser:1983yg}. In this regime, the dependence of physical observables on the sea-quark masses predicted by CHPT has been shown to be in excellent agreement with lattice QCD simulations performed in the two-flavor sector with optimal domain-wall fermions~\cite{Chiu:2011dz}. 
Within this framework, CHPT has been extensively and successfully employed to investigate a wide range of low-energy QCD phenomena, including meson-meson scattering~\cite{Bijnens:1995yn,GomezNicola:2002tn,DiLuzio:2022gsc}, 
axion physics~\cite{Vonk:2020zfh,Balkin:2020dsr,Gao:2024vkw,Cornella:2023kjq}, quark condensates~\cite{Adhikari:2020ufo,Adhikari:2020kdn,Andersen:2012zc,Adhikari:2020qda}, and the thermodynamic properties of QCD matter at finite isospin density~\cite{Son:2000xc,Carignano:2016lxe,Adhikari:2019zaj,Adhikari:2019mdk}. It has also provided valuable insights into the properties of the QCD $\theta$-vacuum and its associated topological features. 
In particular, earlier studies within CHPT successfully evaluated the topological susceptibility and the leading non-Gaussian cumulant $b_2$ at zero temperature, and were later extended to finite temperatures to explore the temperature dependence of $\chi_t$~\cite{GrillidiCortona:2015jxo}. 
These investigations already incorporated isospin-breaking corrections arising from the light quark mass difference, providing important benchmarks for understanding the structure of the $\theta$-vacuum at low energies. Nevertheless, systematic analyses that simultaneously address the temperature evolution of higher-order topological cumulants and the thermal modification of domain wall properties, while explicitly comparing the isospin-symmetric and isospin-breaking limits within a single unified framework, have remained largely unexplored.

The present study aims to fill this gap by extending the SU(2) CHPT analysis to finite temperature and by providing a coherent description of how topological observables and $\theta$-vacuum domain wall dynamics evolve with thermal effects and isospin asymmetry. 
First, we derive the general ground-state solution for arbitrary vacuum phase, thereby going beyond small-$\theta$ expansions~\cite{Mao:2009sy,Bernard:2012ci} and establishing a framework that unifies the treatment of different physical observables. Second, we present a systematic comparison between the isospin-symmetric limit and the realistic isospin-breaking case, thereby clarifying how quark mass asymmetry impacts the temperature evolution of topological observables. The isospin-symmetric limit serves as a theoretically clean and widely used benchmark, particularly in lattice QCD studies where the up and down quark masses are taken equal and electromagnetic effects are neglected. Calculations in this limit enable direct comparison with lattice results that are typically obtained under isospin symmetry and help disentangle genuine thermal effects from those induced by quark mass splitting. In the present work, we analyze both the isospin-symmetric and isospin-breaking cases to quantify how the breaking of isospin symmetry modifies the temperature dependence of topological observables relative to this well-defined baseline. Third, we compute not only the topological susceptibility but also the normalized fourth- and sixth-order cumulants and the domain wall tension as functions of the temperature. Our results confirm that $\chi_t$ agrees with lattice data at low $T$ but deviates at higher $T$, where CHPT is expected to break down. More importantly, we show that $b_{2}$ and the domain wall tension decrease monotonically with temperature, while $b_{4}$ increases, thus establishing the qualitative thermal behavior of the hierarchy of cumulants and the corresponding vacuum domain wall properties in the chiral effective field theory. 
Taken together, these results extend and integrate earlier CHPT studies by demonstrating how isospin breaking and higher-order fluctuations jointly shape the $\theta$-vacuum structure at finite temperature. By providing a unified treatment of susceptibility, higher cumulants, and domain wall tension in both the isospin limit and the broken case, our work delivers new insights into the dynamics of QCD topology and offers valuable theoretical input for axion phenomenology and the physics of topological defects in hot QCD matter.

The remainder of this article is structured as follows. In Sec.~\ref{sec:thetavacuum}, we derive the general ground-state solution of the QCD $\theta$-vacuum for an arbitrary vacuum phase within SU(2) CHPT and obtain the corresponding vacuum free energy density at finite temperature up to next-to-leading order (NLO). This provides the theoretical foundation for the subsequent analysis. In Sec.~\ref{sec:vacuumPro}, we compute the topological susceptibility and evaluate the thermal evolution of the normalized fourth- and sixth-order cumulants of the topological charge distribution, systematically comparing the results in both the isospin limit and with explicit isospin breaking. These results extend previous studies that were mostly restricted to the susceptibility or to zero-temperature cumulants. Sec.~\ref{sec:domainWall} then explores the behavior of the domain wall tension as a function of temperature. 
Finally, in Sec.~\ref{sec:CONCLUSION} we present our conclusions and discuss potential directions for future work.

%%%%%%%%%%%%%%%%%%%%%%%%%%%
\section{Vacuum energy at finite temperature} \label{sec:thetavacuum}

%%%%%%%%%%%%%%%%%%%%%%%%%%%%%%%%%
\subsection{Leading order}

To study physical phenomena related to the $\theta$ parameter, it is usually necessary to eliminate the $\theta$-term from the QCD Lagrangian. This can be achieved by a chiral rotation of the quark fields, leaving the $\theta$ parameter only in the terms involving the quark mass matrix. In the low-energy regime, the relevant physical degrees of freedom are described by pseudoscalar meson fields. Consequently, the resulting Lagrangian can be further matched to the chiral Lagrangian \cite{Gasser:1983yg}. 
The leading order $\mathcal{O}(p^2)$ chiral Lagrangian of SU(2) CHPT including the $\theta$ parameter, which determines the $\theta$-vacuum energy density, is expressed as
\begin{align}\label{eq:Lp2}
\mathcal{L}_{p^2}^{\text{(tree)}}
=
\frac{F^2}{4}\left\langle \chi_\theta U^{\dagger}+ U\chi_\theta^{\dagger}\right\rangle ,
\end{align}
where $F$ denotes the pion decay constant, and the term $\chi_{\theta}$ is defined as $\chi_{\theta} = 2B\mathcal{M}_q \exp(i\mathcal{X}\theta)$, with $B$ being a low-energy constant related to the quark condensate in the chiral limit. The diagonal quark mass matrix is given by $\mathcal{M}_q=\mathrm{diag}\{m_u,m_d\}$, and the phase factor $\mathcal{X}$ takes the diagonal form 
\begin{eqnarray}\label{eq:Xa}
\mathcal{X}=\mathrm{diag}\{\mathcal{X}_u,\mathcal{X}_d\},
\end{eqnarray}
and satisfies the unitarity condition $\langle\mathcal{X} \rangle=1$. 
In this work, we focus on the-flavor case with $N=2$, for which the degrees of freedom of the system are the isospin triplets of the pion mesons. However, as the main focus of this work is on the properties of the $\theta$-vacuum, these Goldstone boson fields are neglected. 
Without loss of generality, the vacuum state of the system can be parameterized as the following diagonal form
\begin{align}\label{U0exp}
U=\left( \begin{array}{cc} e^{i\varphi} & 0 \\ 0 & e^{-i\varphi} \\ \end{array}
\right). 
\end{align}
Then, by substituting Eq.~(\ref{U0exp}) into Eq.~(\ref{eq:Lp2}), we immediately obtain 
\begin{align} \label{eq:Lp2Va}
\mathcal{L}_{p^2}^{\text{(tree)}}=  F^2B
(m_u\cos\phi_u 
+m_d\cos\phi_d),
\end{align}
where $\phi_u$ and $\phi_d$ are defined as
\begin{align}
\begin{cases}
\phi_u= \mathcal{X}_u\theta-\varphi ,\cr
\phi_d= \mathcal{X}_d\theta+\varphi .
\end{cases}
\label{eqPhiDefi}
\end{align}
To determine the ground state, we can take the first derivative of the vacuum energy in Eq.~(\ref{eq:Lp2Va}) with respect to the angle $\varphi$, which yields
\begin{align} \label{Stationary}
z\sin\phi_u=
\sin\phi_d , 
\end{align}
where $z=m_u/m_d$. In contrast to the SU(3) case~\cite{Lu:2020rhp}, where analytical solution is impossible to obtain, we can find an explicit analytical solution for the vacuum state here, i.e.,
\begin{align} \label{phiSolu}
&&
\varphi=\arctan\left[\frac{z\sin \left(\mathcal{X}_u\theta\right)-\sin \left(\mathcal{X}_d\theta\right)}{z\cos \left(\mathcal{X}_u\theta\right)+\cos \left(\mathcal{X}_d\theta\right)}\right].
\end{align}
It is clear that Eq.~(\ref{Stationary}) and the associated physical results depend on the linear combination $\phi_f$, rather than on $\mathcal{X}_f$ and $\varphi$ separately. This implies that $\phi_f$ is the physically meaningful quantity, while $\mathcal{X}_f$ and $\varphi$ are not. 
For example, it is evident that if we assign certain values to $\mathcal{X}_u$ and $\mathcal{X}_d$ with $\mathcal{X}_u=\mathcal{X}_d=1/N$, the above ground state solution reduces to 
\begin{align} \label{eq:phiSoluPLB}
\varphi=\arctan\left[
\frac{1-z}{1+z}
\tan\left(\frac{\theta}{2}\right)\right].
\end{align}
This result recovers the one derived in Ref.~\cite{Guo:2015oxa}. However, we should emphasize that an arbitrary choice of $\mathcal{X}$ in Eq.~(\ref{eq:Xa}) may lead to the mixing of meson fields appearing in the leading order Lagrangian, which complicates the relevant dynamical calculations.
In this case, 
the phase factor $\mathcal{X}$ can also be chosen as 
$
\mathcal{X} = \mathcal{M}_q^{-1}/\langle \mathcal{M}_q^{-1}\rangle  
~ \text{with}~\varphi = 0$~\cite{Georgi:1986df,Kim:1986ax}. 
This parametrization is also the standard starting point for extending the present $\theta$-vacuum analysis to axion-meson interactions, since it matches the chiral Lagrangian to the axion effective field theory without tree-level axion-meson mixing~\cite{GrillidiCortona:2015jxo,Lu:2025fke,Wang:2023xny}. 
Substituting the solution given by Eq.~(\ref{phiSolu}) or Eq.~(\ref{eq:phiSoluPLB}) into Eq.~(\ref{eq:Lp2Va}), one readily obtains the leading-order $\mathcal{O}(p^2)$ chiral Lagrangian, as shown in Eq.~(\ref{eq:Lp2}) and originally derived in Ref.~\cite{DiVecchia:1980yfw},
\begin{align}  \label{eq:Lp2LO}
\begin{aligned}
\mathcal{L}_{p^2}^{\text{(tree)}}
=  F^2M_\pi^2(\theta) ,
\end{aligned}
\end{align}
with the leading order pion mass squared in a $\theta$-vacuum,  $M_\pi^2(\theta)$, given by 
\begin{align}
    M_\pi^2(\theta)=M_\pi^2 \sqrt{1-\frac{4z}{(1+z)^2} \sin ^{2} \left(\frac{\theta}{2}\right)} , 
\end{align}
where $M_\pi^2=Bm_d(1+z)$ is the leading order pion mass squared in the vacuum.

%%%%%%%%%%%%%%%%%%%%%%%%%%%%%%%%%
\subsection{Next-to-leading order}

To obtain precise theoretical predictions on the low energy physics of the $\theta$-vacuum up to NLO, we must further consider the contributions at $\mathcal{O}(p^4)$, including both tree-level and one-loop diagram contributions. The chiral Lagrangian expression for the tree-level terms relevant to calculating the $\theta$-vacuum free energy density at $\mathcal{O}(p^4)$ is given by
\begin{equation} \label{eq:Lp4tree}
\begin{aligned}
\mathcal{L}_{p^4}^{\text{(tree)}}
=& \frac{l_3}{16}\left\langle\chi_{\theta} U^{\dagger}+ U\chi_{\theta}^{\dagger}\right\rangle^2-\frac{l_7}{16}\left\langle\chi_{\theta} U^{\dagger}-U\chi_{\theta}^{\dagger}\right\rangle^2 \\% \nonumber\\
& +\frac{h_1+h_3}{4}\left\langle\chi_{\theta}^{\dagger} \chi_{\theta}\right\rangle+\frac{h_1-h_3}{2} \operatorname{Re}\left(\operatorname{det} \chi_{\theta}\right).
\end{aligned}
\end{equation} 
Note that the low energy constants $l_7$ and $h_3$ are scale independent, whereas the low energy constants $l_3$ and $h_1$ contain both finite and divergent parts, which are necessary to cancel out the divergent terms that appear in higher order expressions. The relationship between the bare and the renormalized constants are given by~\cite{Gasser:1983yg}
\begin{align}
l_{3}= l_{3}^{r}-\frac{\lambda}{2},~~~~~~
h_1=h_1^r+2\lambda, 
\end{align}
where $\lambda$ denotes the divergent factor introduced in the dimensional regularization when the spacetime dimension $d$ is set to 4, namely
\begin{align}
\lambda=\frac{\mu^{d-4}}{16 \pi^{2}}\left\{\frac{1}{d-4}-\frac{1}{2}\left[\ln (4 \pi)+\Gamma^{\prime}(1)+1\right]\right\},
\end{align}
where $\mu$ represents the scale of the dimensional regularization.
In addition, using the dimensional regularization, we can compute the one-loop contribution to the vacuum energy density at the NLO~\cite{Guo:2015oxa}, which gives
\begin{align} \label{eq:Vloop}
\begin{aligned}
\mathcal{L}_{p^4}^{\text{(loop)}}
=&~ \frac{3}{2} 
\int i\frac{d^dp}{(2\pi)^d} \ln \left[-p^2 +M_\pi^2(\theta) \right] \\
=&~ 
\frac{3M_\pi^4(\theta)}{128\pi^2}\left[ 
1-2\ln\frac{M_\pi^2(\theta)}{\mu^2}\right]+ \mathcal{L}_{p^4}^{\text{(div)}} , 
\end{aligned}
\end{align}
where $\mathcal{L}_{p^4}^{\text{(div)}}=-3\lambda M_\pi^4(\theta)/2$ represents the divergence arising from the one-loop contributions, and the term proportional to $\lambda$ collects all ultraviolet divergent pieces at this order.

%%%%%%%%%%%%%%%

In the framework of finite-temperature field theory, the incorporation of temperature effects is achieved through the replacement of the zero component of the four-dimensional momentum, $p_0$, by a discrete sum over Matsubara frequencies, namely $p_0=i\omega_n=i2n\pi T$ for bosonic fields. This substitution naturally embeds thermal excitations into the theoretical formalism and plays a central role in the evaluation of loop diagrams at nonzero temperature. In particular, the finite-temperature corrections to the vacuum energy density can be systematically derived by performing this replacement in the one-loop functional determinants. 
At zero temperature, the free energy density in SU(2) CHPT up to NLO includes contributions from the tree-level term, the NLO contact interactions, and the one-loop diagrams, corresponding respectively to Eqs.~(\ref{eq:Lp2LO}), (\ref{eq:Lp4tree}), and (\ref{eq:Vloop}). Summing these pieces yields the complete expression for the vacuum free energy density at vanishing temperature, given as
\begin{align}
V(\theta)
=& -M_\pi^2(\theta) F_\pi^2  \Bigg\{1-2 \frac{M_\pi^2}{F_\pi^2}\bigg[-\frac{\left(1-z\right)^2}{\left(1+z\right)^2} l_7^r\nonumber\\
&+l_3^r+l_4^r-\frac{3}{64 \pi^2} \log \left(\frac{M_\pi^2}{\mu^2}\right)\bigg] %\nonumber\\
+\frac{M_\pi^2(\theta)}{F_\pi^2} \nonumber\\
& \times \Bigg[\frac{4 z^2}{\left(1+z\right)^4} \frac{M_\pi^8 \sin ^2(\theta)}{M_\pi^8(\theta)} l_7^r+(h_1^r-h_3^r+l_3^r) \nonumber\\
& -\frac{3}{64 \pi^2}\Bigg(\log \left(\frac{M_\pi^2(\theta)}{\mu^2}\right)-\frac{1}{2}\Bigg)\Bigg]\Bigg\}.
\end{align}
In the above, the pion decay constant in the chiral limit, $F$, has been replaced with the physical value $F_\pi$, ensuring that higher-order effects are consistently absorbed into renormalized low-energy constants.

Within the framework of CHPT, thermal effects have been systematically incorporated to explore the properties and dynamical behavior of light scalar resonances~\cite{Gao:2019idb,Gu:2018swy}, axion-meson scattering~\cite{GomezNicola:2002tn,Wang:2023xny,GomezNicola:2023ghi}, as well as scalar susceptibilities and four-quark condensates~\cite{GomezNicola:2012uc} at finite temperature.
In addition, the topological susceptibility has been computed up to NNLO in the context of U(3) CHPT at finite temperature~\cite{GomezNicola:2019myi}. 
The corresponding free energy density, evaluated in our work in SU(2) CHPT up to NLO at finite temperature, finally can be expressed as %~\cite{Smilga:1995qf}
\begin{align} \label{eq:vthetaT}
V(\theta,T)= &V(\theta)+ 
\frac{3T}{2\pi^2}\int_0^\infty  p^2 \nonumber\\
&\times \log\left[1-\exp\left(-\frac{\sqrt{p^2+M_\pi^2(\theta)}}{T}\right)\right]dp,
\end{align}
where $V(\theta) = V(\theta, T=0)$ denotes the zero-temperature contribution to the $\theta$-vacuum energy density. As the temperature $T$ approaches zero, $V(\theta, T)$ naturally reduces to this zero-temperature form, $V(\theta)$. 
It can be verified that the ultraviolet divergent contributions arising from the loop integrals are precisely canceled by those contained in the $\mathcal{O}(p^4)$ counterterms of the effective Lagrangian. This cancellation demonstrates the self-consistency of CHPT and ensures that the final expression for the vacuum energy density is finite and renormalization-scale independent. In this way, one recovers a physically meaningful quantity in line with the general principle that all observables must be finite.

%%%%%%%%%%%%%%%%%%%%%%%%%%
\section{Topological observables at finite temperature}
\label{sec:vacuumPro}

The distribution of the QCD topological charge can be described by cumulants, of which the topological susceptibility is the lowest order. In a $\theta$-vacuum, the vacuum energy density serves as the generating function for these cumulants. These topological quantities play a crucial role in understanding the QCD vacuum structure and in extracting physical observables from lattice simulations with fixed topology~\cite{Aoki:2007ka,Bautista:2015yza}. 
Various methods exist to measure these quantities in the lattice framework. However, accurate results from lattice simulations of the topological susceptibility are currently only available for zero and high temperature regions~\cite{Athenodorou:2022aay,Bonati:2018blm}. In contrast, accurate calculations near and below the QCD phase transition temperature are still lacking. Bridging this gap is essential to advance our understanding of non-perturbative QCD phenomena and to refine lattice QCD predictions in the physically relevant temperature range.

The topological susceptibility, which is the lowest order cumulant of the QCD topological charge distribution, is defined as 
\begin{align} \label{eq:chit}
\chi_t=\frac{\left\langle Q^2\right\rangle_{\theta=0}}{\mathcal{V}}=\frac{d^2}{d\theta^2}V(\theta,T)\bigg|_{\theta=0} ,
\end{align}
where $\mathcal{V}$ denotes the space-time volume, and $Q$ is the topological charge defined in Eq.~(\ref{eq:Qcharge}). At zero temperature, the topological susceptibility computed in SU(2) CHPT up to NLO is estimated to be $\chi_t^{1/4}=75.5(5)$ MeV~\cite{GrillidiCortona:2015jxo}, which is in excellent agreement with the lattice simulation result of $\chi_t^{1/4}=75.6(2)$ MeV~\cite{Borsanyi:2016ksw}, taking into account the isospin breaking effects due to the light quark mass difference. While in the isospin limit case, the CHPT prediction and lattice simulation also agree with each other, which give $\chi_t^{1/4}=77.9(2)$ MeV and $\chi_t^{1/4}=78.1(2)$ MeV, respectively. 

\begin{figure}[bt]
\centering
  \includegraphics[width=0.48\textwidth]{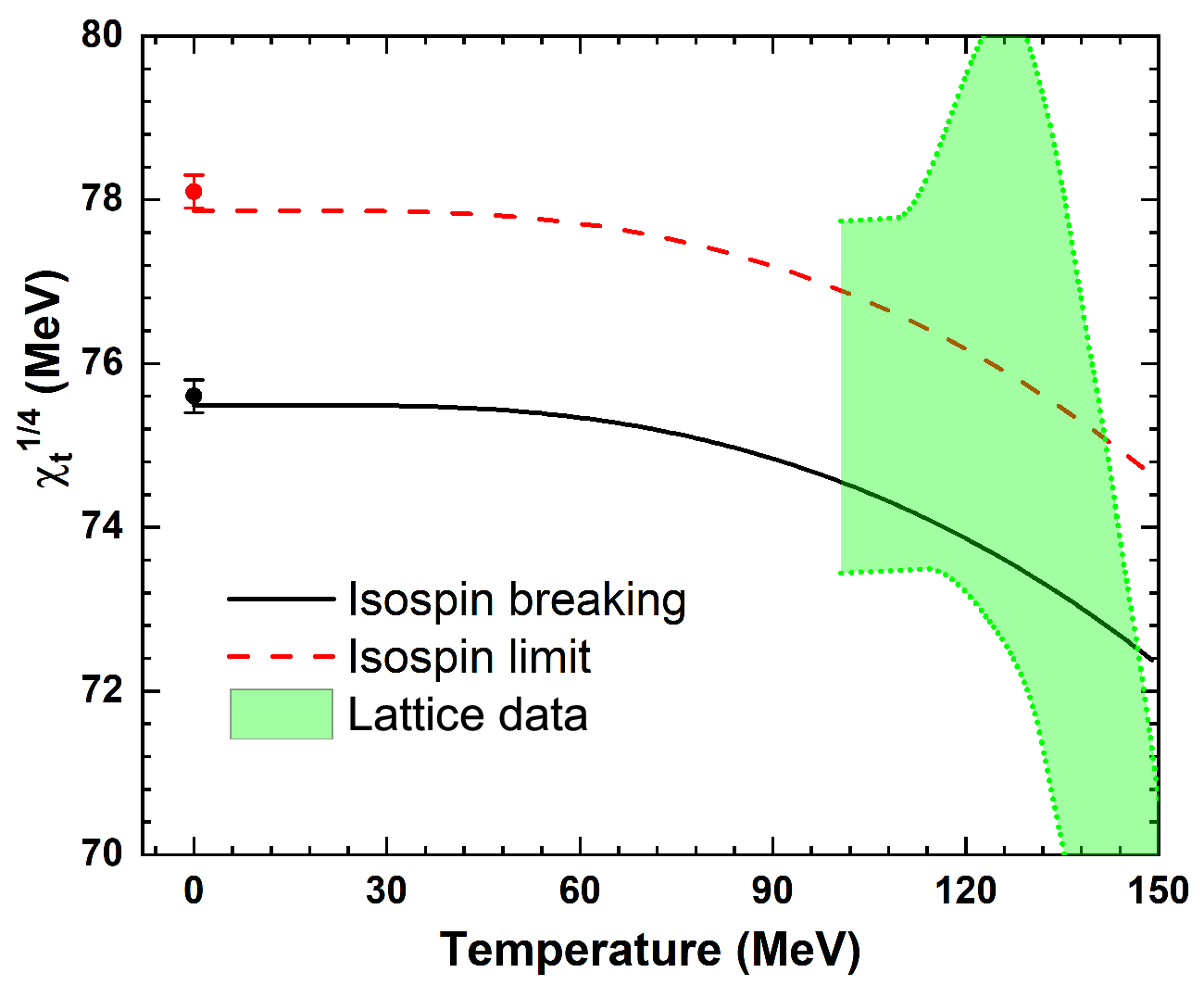}\\
  \caption{The temperature dependence of the topological susceptibility, $\chi_t^{1/4}$, calculated within CHPT. For comparison, lattice QCD results from Ref.~\cite{Borsanyi:2016ksw} are shown as the green band at finite temperature, with the zero-temperature value indicated by the data point with error bar.
 }\label{fig:xT}
\end{figure}

To explore the temperature dependence of the QCD topological susceptibility, Fig.~\ref{fig:xT} displays $\chi_t^{1/4}$ as a function of the temperature. The black solid curve corresponds to the isospin-breaking case with $m_u \neq m_d$, while the red dashed curve denotes the isospin-symmetric limit with $m_u = m_d$. For comparison, the available lattice QCD results at low temperatures, taken from Ref.~\cite{Borsanyi:2016ksw}, are represented by the green band. As the temperature increases, the predictions from CHPT exhibit a smooth monotonic decrease of $\chi_t^{1/4}$ in both the isospin-breaking and isospin-symmetric cases, reflecting the progressive suppression of topological fluctuations in the thermal medium. The isospin-breaking effect manifests itself in a slightly lower magnitude of $\chi_t^{1/4}$ across the entire temperature range, in agreement with the expectation that mass splitting between light quarks tends to reduce the overall topological response.

In the low-temperature region, where lattice data are available, our CHPT results reproduce the lattice trend with good quantitative agreement within uncertainties. This consistency provides support for the reliability of CHPT in capturing the nonperturbative topological features of the QCD vacuum up to temperatures of order $T \lesssim 150~\mathrm{MeV}$. At higher temperatures, where the lattice data become sparse and the effective field theory approach is expected to gradually break down, our results can nevertheless serve as a controlled extrapolation, offering complementary insights into the interplay between isospin effects and thermal suppression of topological susceptibility.

For the numerical calculations presented in this work and in the subsequent sections, we adopt as inputs the pion decay constant and mass, $F_\pi = 92.2~\mathrm{MeV}$ and $M_\pi = 134.98~\mathrm{MeV}$, as reported by the Particle Data Group~\cite{ParticleDataGroup:2018ovx}.
The light-quark mass ratio is taken to be $z = m_u/m_d = 0.48 \pm 0.03$, consistent with recent lattice QCD determinations~\cite{FlavourLatticeAveragingGroup:2019iem}.
In addition, the low-energy constants $l_7 = (7 \pm 4)\times 10^{-3}$ and $h_1^r - h_3 - l_4^r = (4.8 \pm 1.4)\times 10^{-3}$ are adopted from Ref.~\cite{GrillidiCortona:2015jxo}.
By construction, these parameters are independent of both temperature and renormalization scale within the CHPT framework~\cite{Fernandez-Fraile:2009eug}.

\begin{figure}[bt]
\centering
  \includegraphics[width=0.48\textwidth]{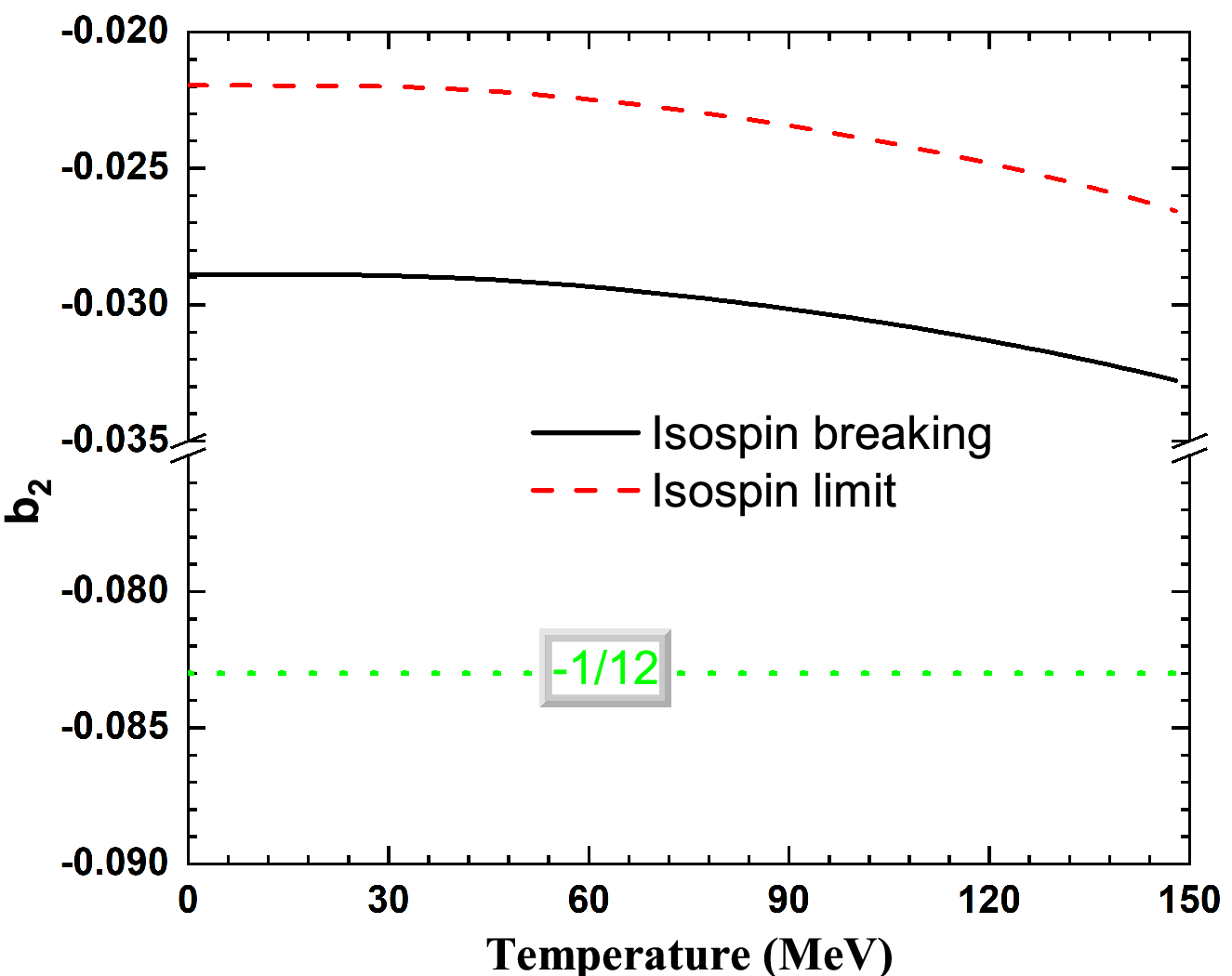}\\
  \caption{The normalized fourth-order cumulant of the QCD topological charge distribution as a function of the temperature, with (solid curve) and without (dashed curve) isospin breaking. The green curve denotes the asymptotic value $b_2^{\text{inst}}$ predicted by the dilute instanton gas model~\cite{Bonati:2015uga}.
 }\label{fig:b2T}
\end{figure}

To quantify deviations of the topological charge distribution from a purely Gaussian behavior, it is customary to analyze the higher-order normalized cumulants of the distribution, which systematically characterize the non-Gaussian features of the QCD vacuum. In particular, the normalized fourth-order cumulant, $b_2$, measures the leading deviation from Gaussianity and provides direct information on the shape of the topological charge distribution, while the normalized sixth-order cumulant, $b_4$, captures subleading corrections and quantifies the degree of asymmetry or kurtosis beyond the leading order. These cumulants serve as essential observables for testing the validity of effective models such as CHPT and the Nambu$-$Jona-Lasinio (NJL) model against lattice QCD data, especially at finite temperature where topological fluctuations are expected to be suppressed.

Formally, the normalized cumulants $b_2$ and $b_4$ are defined as~\cite{Bonati:2015sqt}
\begin{align}
b_2=-\frac{\left\langle Q^4\right\rangle_{\theta=0}-3\left\langle Q^2\right\rangle_{\theta=0}^2}{12\left\langle Q^2\right\rangle_{\theta=0}}
=-\frac{c_4}{12\chi_t} ,
\end{align}
and 
\begin{align}
    b_4 =\frac{\left[\left\langle Q^6\right\rangle-15\left\langle Q^2\right\rangle\left\langle Q^4\right\rangle+30\left\langle Q^2\right\rangle^3\right]_{\theta=0}}{360\left\langle Q^2\right\rangle_{\theta=0}} %\nonumber\\
    =\frac{c_6}{360\chi_t}, \nonumber\\
\end{align}
where $Q$ denotes the topological charge defined in Eq.~(\ref{eq:Qcharge}), $\chi_t$ represents the topological susceptibility given in Eq.~(\ref{eq:chit}), and the coefficients $c_{2n}$ correspond to the even-order derivatives of the vacuum free energy density $V(\theta, T)$ with respect to the vacuum angle:
\begin{align}
c_{2n} = \frac{d^{2n}}{d\theta^{2n}}V(\theta,T)\bigg|_{\theta=0}.
\end{align}

The cumulants $b_2$ and $b_4$ thus encode complementary information about the curvature and higher-order structure of the $\theta$-vacuum potential. A purely Gaussian distribution would yield $b_2 = b_4 = 0$, whereas any nonzero values directly reflect the presence of higher-order topological correlations in the QCD vacuum. In lattice QCD and effective field theory studies, $b_2$ typically takes a small negative value, implying that the topological charge distribution is more sharply peaked than a Gaussian one, while $b_4$ often exhibits larger uncertainties due to the limited sensitivity of current simulations. The temperature dependence of these quantities provides a sensitive probe of the transition from the nonperturbative topological regime to the perturbative high-temperature phase, making them key observables for understanding the microscopic dynamics of the $\theta$-vacuum and its role in axion physics.

In Fig.~\ref{fig:b2T}, we show the temperature dependence of the normalized fourth-order cumulant $b_2$ of the QCD topological charge distribution. The black solid curve corresponds to the isospin-breaking case, while the red dashed curve represents the isospin-symmetric case.
It is evident that in both cases $b_2$ remains negative throughout the considered temperature range, implying that the topological charge distribution is always more sharply peaked than a Gaussian. As the temperature increases, the absolute value of $b_2$ shows a slight growth, consistent with lattice simulations~\cite{Bonati:2015vqz}, and the effect becomes more pronounced when isospin symmetry is broken. This behavior suggests that deviations from Gaussianity are further enhanced at higher temperatures in the presence of isospin breaking. 
Notably, the overall tendency of $|b_2|$ to increase with rising temperature is consistent with results from lattice QCD and the NJL model~\cite{Lu:2025bpd}, both of which predict that, in the high-temperature limit, $b_2$ converges to the asymptotic value expected from the dilute instanton gas model, namely $b_2^{\text{inst}}=-1/12 \simeq-0.083$~\cite{Bonati:2015uga,Gong:2024cwc}.

\begin{figure}[bt]
\centering
%  \includegraphics[width=0.48\textwidth]{b2T.eps}\\
  \includegraphics[width=0.49\textwidth]{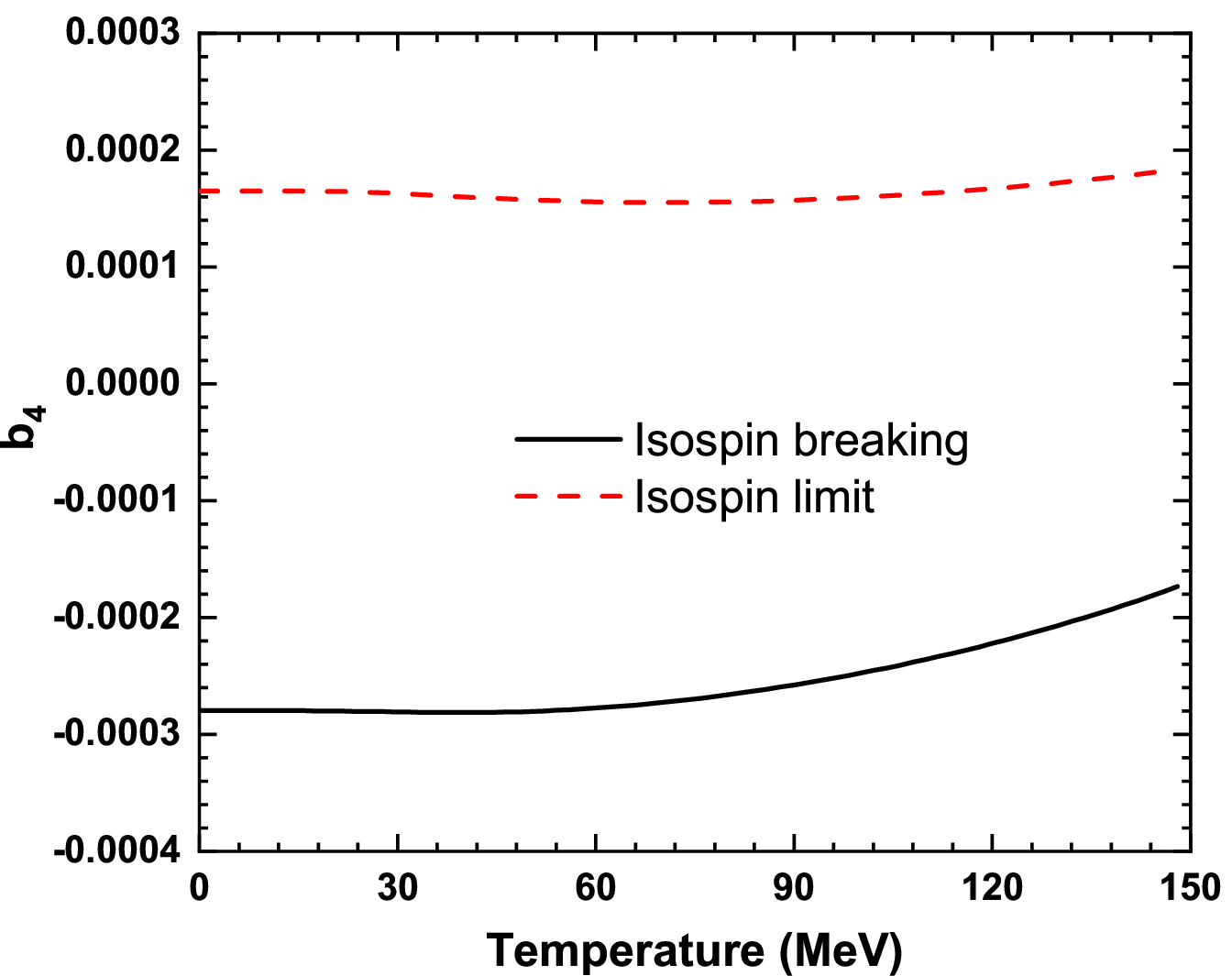}\\
  \caption{The normalized sixth-order cumulant of the QCD topological charge distribution as a function of the temperature, with (solid curve) and without (dashed curve) isospin breaking.
 }\label{fig:b4T}
\end{figure}

To complement the analysis of the normalized fourth-order cumulant, Fig.~\ref{fig:b4T} shows the temperature dependence of the normalized sixth-order cumulant $b_4$ of the QCD topological charge distribution. The black solid curve represents the isospin-breaking case with $m_u \neq m_d$, while the red dashed curve corresponds to the isospin-symmetric case with $m_u = m_d$. 
We observe that $b_4$ exhibits markedly different behaviors in the two cases. In the isospin-symmetric limit, $b_4$ remains positive and nearly constant across the entire temperature range, suggesting only mild temperature sensitivity. In contrast, once isospin breaking is introduced, $b_4$ turns negative and shows a slow increase with temperature, approaching zero around $T \simeq 150$ MeV. This qualitative change highlights the strong impact of isospin breaking on higher-order cumulants, and indicates that non-Gaussian corrections to the $\theta$-dependent effective potential are more significant in this case. Together with the results for $b_2$, this behavior underscores the role of higher-order cumulants as sensitive probes of the QCD $\theta$-vacuum structure under varying thermodynamic conditions.

%%%%%%%%%%%%%%%%%%%%%%%%%%%%%%%%%%%%%%

\section{$\theta$-dependent effective potential and domain wall tension}
\label{sec:domainWall}

The energy per unit area associated with domain walls in the QCD $\theta$-vacuum, known as the domain wall tension $\sigma$, can be calculated using the following formula
\begin{align} \label{eq:sigmasigma0}
\sigma = 2\sqrt{2} f_a \int_0^\pi d\theta \sqrt{\Delta  V(\theta,T)},
\end{align} 
where $\Delta V(\theta,T)= V(\theta,T)-V(0,T)$ denotes the temperature-dependent deviation of the effective potential of the QCD $\theta$-vacuum from that at $\theta=0$. In the axion effective theory limit where the axion field tracks the QCD angle, $a/f_a\simeq \theta$, the integral in Eq.~(\ref{eq:sigmasigma0}) also yields the axion domain wall tension. Our result for $\sigma$ at finite temperature therefore quantifies how thermal effects soften the barrier separating adjacent vacua, informing the energetics and evolution of axion wall networks.

\begin{figure}[bt]
\centering
  \includegraphics[width=0.48\textwidth]{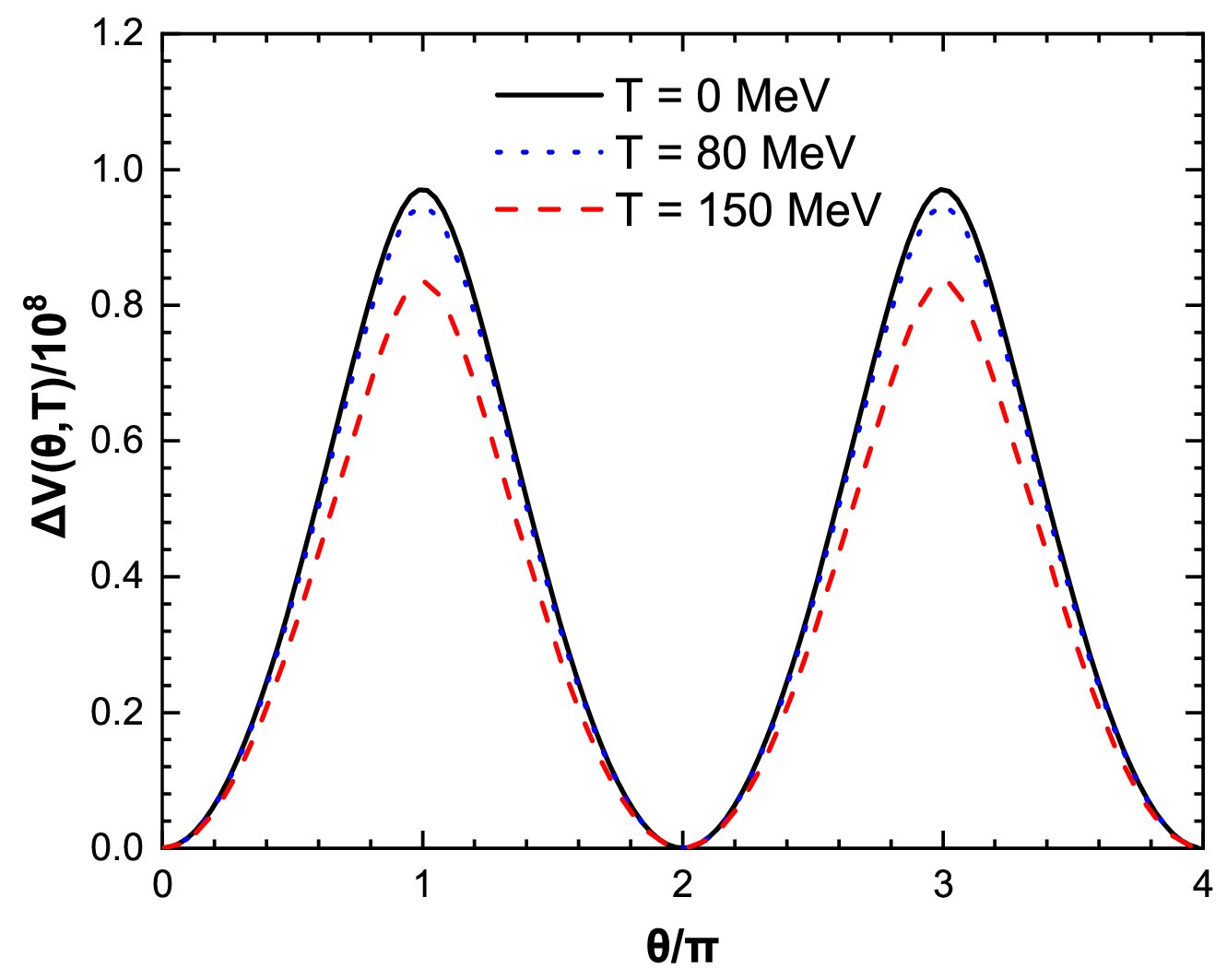}\\
  \caption{Free energy density of the $\theta$-vacuum, $\Delta V(\theta, T)$, as a function of the angle $\theta$ for different temperatures.
 }\label{fig:vtheta}
\end{figure}

In Fig.~\ref{fig:vtheta}, the free energy density of the $\theta$-vacuum, $\Delta V(\theta, T)$, is shown for the isospin-breaking case as a function of the vacuum angle $\theta$ at several representative temperatures. 
At $T=0$ (black solid curve), the potential exhibits the expected periodicity in $\theta$, with minima located at integer multiples of $\pi$, reflecting the discrete symmetry of the QCD vacuum. As the temperature increases to $T=80$~MeV (blue dotted curve) and further to $T=150$~MeV (red dashed curve), the overall magnitude of the potential is reduced, signaling the thermal suppression of topological fluctuations. Importantly, while the periodic structure of the $\theta$ dependence remains preserved, the height of the potential barriers between adjacent minima decreases systematically with increasing temperature. This behavior indicates a smoother interpolation among distinct $\theta$-vacua at finite temperature. Moreover, the observed reduction in the barrier height with temperature is qualitatively consistent with recent results obtained within the NJL model, which likewise predict a suppression of the extremal values of $\Delta V(\theta,T)$ as the system is heated~\cite{Lu:2018ukl}.

\begin{figure}[bt]
\centering
  \includegraphics[width=0.48\textwidth]{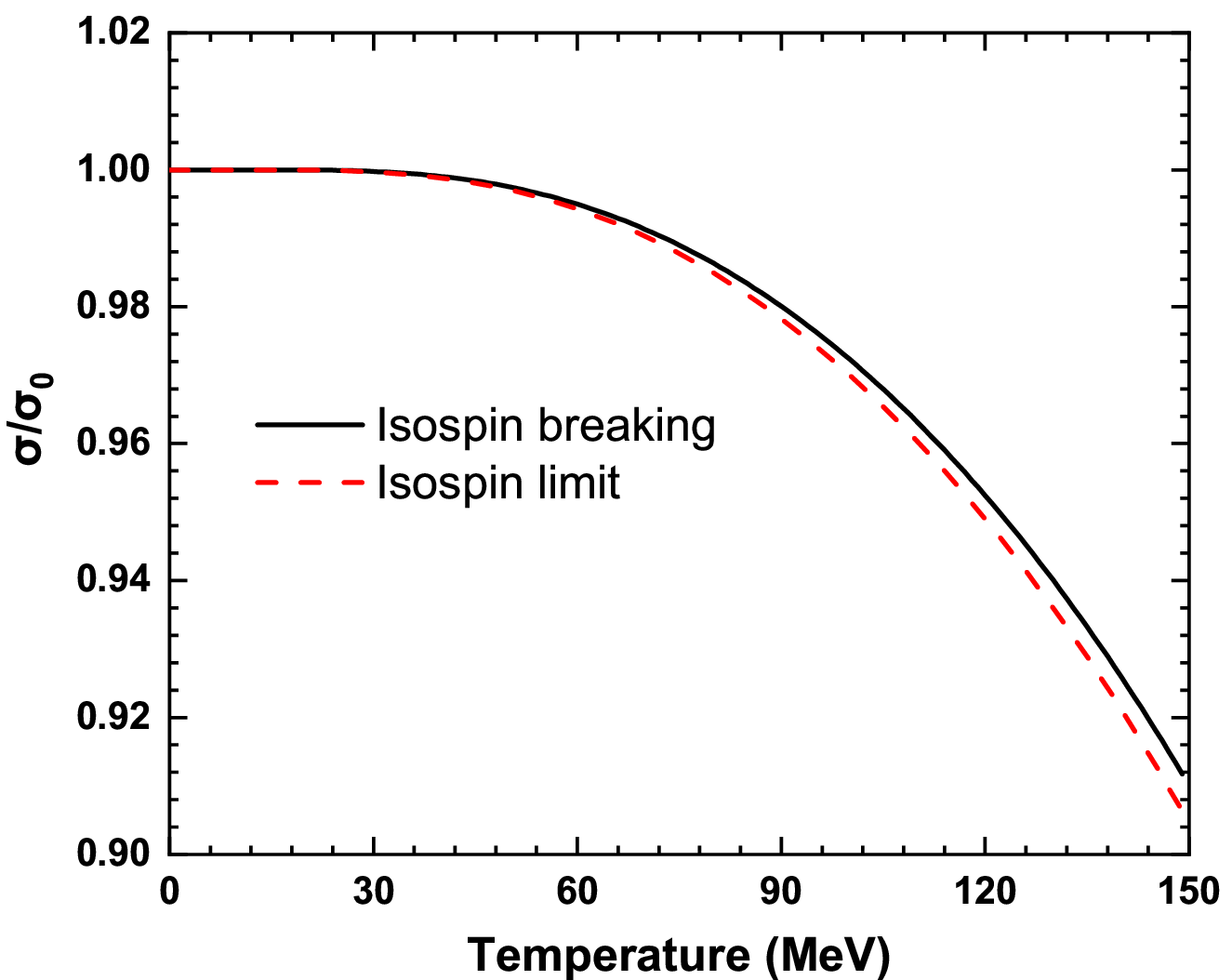}\\
  \caption{Normalized domain wall tension, $\sigma/\sigma_0$, of the QCD $\theta$-vacuum as a function of temperature, calculated with (solid curve) and without (dashed curve) isospin breaking.
 }\label{fig:tensionT}
\end{figure}

Figure~\ref{fig:tensionT} shows the temperature dependence of the normalized domain wall tension, $\sigma/\sigma_0$, in the QCD $\theta$-vacuum, where $\sigma_0$ denotes the zero-temperature value. The solid black curve corresponds to the isospin-breaking case with $m_u \neq m_d$, while the red dashed curve represents the isospin-symmetric limit with $m_u = m_d$. At very low temperatures, $T \lesssim 40~\mathrm{MeV}$, the normalized domain wall tension remains nearly constant, indicating that thermal excitations have a negligible impact on the stability of the $\theta$-vacuum. As the temperature increases toward the intermediate region, $T \approx 80~\mathrm{MeV}$, $\sigma/\sigma_0$ begins to decrease gradually, reflecting the onset of thermal suppression of the potential barrier separating neighboring topological vacua. Beyond this point, in the vicinity of the chiral crossover, $T \sim 120-150~\mathrm{MeV}$, the reduction becomes more pronounced, with the tension dropping by about 8$-$10\% relative to its zero-temperature value. This behavior suggests that thermal fluctuations tend to soften the domain wall, indicating a gradual reduction in the energy cost associated with topological transitions in hot QCD matter.

The comparison between the two curves reveals that isospin breaking introduces only a mild quantitative correction: it slightly delays the decrease of $\sigma/\sigma_0$ with temperature but does not alter the overall trend. This small shift arises from the mass asymmetry between the up and down quarks, which marginally modifies the curvature of the effective potential near $\theta=\pi$. The qualitative behavior is consistent with recent findings from NJL-model analyses, where the thermal suppression of domain wall tension follows a similar monotonic pattern. Overall, these results confirm that while the stability of $\theta$-vacuum domain walls is primarily governed by thermal effects, isospin breaking plays only a secondary role within the temperature range relevant to the chiral transition.

%------------------------------------
\begin{table}
%\squeezetable
\caption{\label{table:CHPTT0} 
Numerical values of the cumulants of the QCD topological charge distribution at zero temperature, obtained within SU(2) CHPT in both the isospin-symmetric and isospin-breaking cases.}
\setlength{\tabcolsep}{0.9pt}
\renewcommand\arraystretch{1.7}
\begin{ruledtabular}
\vspace{+0.1cm}
\begin{tabular*}{\hsize}{@{}@{\extracolsep{\fill}}cccc@{}}
& $\chi_t^{1/4}$ (MeV)      &  $b_2$      &   $b_4$   \\  
         \hline
Isospin breaking &  75.5(5)     &  -0.029(2)       &  -0.00028(7)     \\
Isospin limit &  77.9(4)     &  -0.022(1)      &  -0.00017(6)     \\
\end{tabular*}
\end{ruledtabular}
\vspace{-0.5cm}
\end{table}
%------------------------------------

%%%%%%%%%%%%%%%%%%%%%%
\section{Conclusions}  
\label{sec:CONCLUSION}

As the low-energy effective theory of QCD, CHPT is capable of accurately describing the low-energy physics in a $\theta$-vacuum. In this work, we have carried out a systematic study of the $\theta$-vacuum in QCD within the framework of SU(2) CHPT at finite temperature up to NLO. By deriving the general ground-state solution for arbitrary vacuum phase, we established a unified expression for the vacuum energy that allows for a consistent evaluation of topological observables beyond small $\theta$ expansions. Building on earlier works that analyzed the susceptibility and low-order cumulants either at $T=0$~\cite{Guo:2015oxa} or at finite $T$~\cite{GrillidiCortona:2015jxo}, we extended the analysis to include higher cumulants and the domain wall tension, providing a more comprehensive characterization of the QCD $\theta$-vacuum.

Our results reveal several novel features. By considering both the isospin-symmetric limit and the realistic isospin-breaking case, we clarified the role of quark mass asymmetry in shaping the thermal behavior of topological observables, thereby addressing an important source of uncertainty for phenomenological applications. We showed that the topological susceptibility $\chi_t$ is consistent with lattice data at low temperatures but deviates as the system approaches the chiral crossover, signaling the breakdown of the chiral expansion. Moreover, we demonstrated that the normalized fourth-order cumulant $b_{2}$ and the domain wall tension decrease monotonically with increasing temperature, while the normalized sixth-order cumulant $b_{4}$ increases monotonically—establishing, for the first time, the qualitative thermal trends of higher cumulants and their connection to the stability of $\theta$-vacuum domain walls. Since the present analysis is based on NLO SU(2) CHPT, quantitative reliability is limited to temperatures well below the chiral crossover ($T \lesssim 150$ MeV). Beyond this range, higher-order effects and possible resonance contributions may become non-negligible. Taken together, these results extend previous CHPT-based analyses and open new perspectives for understanding the interplay between topology, axion-related effective theories, and QCD thermodynamics.

There are several promising directions to build on this work. One avenue is to extend the analysis to finite isospin chemical potential~\cite{Son:2000xc,Carignano:2016rvs}, where the absence of a sign problem allows direct comparison with lattice QCD. Such studies could provide stringent tests of CHPT predictions for $\chi_t$ and higher-order cumulants and help constrain the associated low-energy constants. Another important step is the inclusion of baryon chemical potential and baryonic degrees of freedom~\cite{Springmann:2024mjp,DiLuzio:2024vzg}, which are essential for describing dense matter in compact stars. Incorporating these effects would not only improve the realism of the effective theory but also strengthen the connection between QCD topology, axion phenomenology, and astrophysical applications. For axion-related applications, the temperature-dependent topological susceptibility $\chi_t(T)$, together with the higher-order cumulants $b_2$ and $b_4$ obtained here, provide ready-to-use inputs for the finite-temperature axion mass and potential. Likewise, the extracted $\sigma(T)$ can be applied to simulations of axion domain-wall dynamics, further bridging chiral effective theory with axion cosmology and hot QCD phenomenology.

\section*{Acknowledgments}

Z.~Y.~L. thanks Prof.~Feng-Kun Guo and Prof.~Zhi-Hui Guo for useful discussions. 
This work is supported in part by 
the National Natural Science Foundation of China 
(Grants No.~12205093, No.~12405054, No.~12404240, and No.~12375045), and the Natural Science Foundation of Hunan Province (Grants No.~2026JJ50352 and No.~2026JJ60316, and No.~2024JJ6210). Z.~Z. is supported by the funding from the Chinese Academy of Sciences (Grants No. E25155U1 and No. E329A3M1), and the Institute of High Energy Physics (Grant No. E3545KU2), as well as the National Program on Key Research and Development Project from the Ministry of Science and Technology of China (Grant No. 2021YFA0718500).

\section*{Data availability}
The data that support the findings of this
article are not publicly available. The data are available
from the authors upon reasonable request.

\end{CJK*}

\bibliographystyle{aapmrev4-2}  %%y%%  %% longbibliography
\bibliography{Ref}

%\end{CJK*}
\end{document}